\magnification=\magstep1
\hsize 6.0 true in
\vsize 9.0 true in
\voffset=-.5truein
\pretolerance=10000  
\baselineskip=24truept

\font\tentworm=cmr10 scaled \magstep2
\font\tentwobf=cmbx10 scaled \magstep2

\font\tenonerm=cmr10 scaled \magstep1 
\font\tenonebf=cmbx10 scaled \magstep1

\font\eightrm=cmr8
\font\eightit=cmti8
\font\eightbf=cmbx8
\font\eightsl=cmsl8
\font\sevensy=cmsy7
\font\sevenm=cmmi7

\font\twelverm=cmr12  
\font\twelvebf=cmbx12
\def\subsection #1\par{\noindent {\bf #1} \noindent \rm}

\def\mid {\let\rm=\tenonerm \let\bf=\tenonebf \rm \bf}

\def\para{\par \vskip 12 pt}

\def\head{\let\rm=\tentworm \let\bf=\tentwobf \rm \bf}

\def\heading #1 #2\par{\centerline {\head #1} \smallskip
 \centerline {\head #2} \vskip .15 pt \rm}

\def\eight{\let\rm=\eightrm \let\it=\eightit \let\bf=\eightbf 
\let\sl=\eightsl \let\sy=\sevensy \let\m=\sevenm \rm}

\def\foots{\noindent \eight \baselineskip=10 true pt \noindent \rm}
\def\sexion{\let\rm=\twelverm \let\bf=\twelvebf \rm \bf}

\def\section #1 #2\par{\vskip 20 pt \noindent {\mid #1} \enspace {\mid #2} 
  \para \noindent \rm}

\def\abstract#1\par{\para \foots {\bf Abstract: \enspace}#1 \para}

\def\author#1\par{\centerline {#1} \vskip 0.1 true in \rm}

\def\abstract#1\par{\noindent {\bf Abstract: }#1 \vskip 0.5 true in \rm}

\def\sqr#1#2{{\vcenter{\vbox{\hrule height.#2pt
  \hbox {\vrule width.#2pt height#1pt \kern#1pt
  \vrule width.#2pt}
  \hrule height.#2pt}}}}

\def\n{\noindent}
\def\s{\smallskip}
\def\m{\medskip}
\def\b{\bigskip}
\def\c{\centerline}

\def\gne #1 #2{\ \vphantom{S}^{\raise-0.5pt\hbox{$\scriptstyle #1$}}_
{\raise0.5pt \hbox{$\scriptstyle #2$}}}

\def\ooo #1 #2{\vphantom{S}^{\raise-0.5pt\hbox{$\scriptstyle #1$}}_
{\raise0.5pt \hbox{$\scriptstyle #2$}}}


\line{\hfill IUCAA - 16/96}

\line{\hfill May 1996}
\m
\m

\c{\bf\mid Isothermal spherical  perfect  fluid  model:} 
\c{\bf\mid Uniqueness and  Conformal mapping}
\b
\b
\b
\b
\b 
\b
\b
\c{\bf Naresh Dadhich\footnote{$^* $}{E-mail : naresh@iucaa.ernet.in}}
\c{\bf Inter University  Centre for Astronomy \& Astrophysics}
\c{\bf P.O. Box 4, Pune-411007, India}
\b
\b
\b
\b
\b
\baselineskip=16truept
\c{\bf Abstract}
\s 
\n We prove the theorem: The necessary and sufficient condition  for 
a  spherically  symmetric spacetime to  represent  an  isothermal 
perfect fluid (barotropic equation of state with density  falling 
off  as  inverse  square of the  curvature  radius)  distribution 
without  boundary  is  that it is conformal  to  the  ``minimally'' 
curved  (gravitation only manifesting in tidal  acceleration  and 
being absent in particle trajectory)  spacetime. 
\b
\b
\b
\b
\n PACS numbers : 0420, 9880.

\vfill\eject

\baselineskip=24truept
\n Except  for  Brinkman's theorem [1] conformally  relating  two 
Einstein spaces, only spacetimes conformal to flat spacetime  have 
been considered. The well-known examples of such spacetimes,  that 
are also of great astrophysical and cosmological interest, are the 
Friedman-Robertson-Walker  (FRW)  model of the Universe  and  the 
Schwarzschild   solution  describing  interior  of  a   star   in 
hydrostatic  equilibrium. These  spacetimes  have   distinguishing 
physical properties, like isotropy and homogeneity for the  former 
and  uniform  density  for the latter, but   no  direct 
linkage  of  them with conformal flatness has been shown.
In this note  we  shall 
establish  a  unique association between isothermality  of  fluid 
with   conformal   character   of   spacetime. Such   a   clear 
characterisation of physical behaviour with geometric property is 
rather very rare. 
\b
\n Here  the  base spacetime is of course not flat  but  ``minimally'' 
curved. It is a spherically symmetric spacetime from which  radial 
acceleration has been anulled out but curvature is not zero, which 
menifests  in tidal acceleration for transverse motion. Though  it 
is  not  a solution of the Einstein equation, but it  presents  an 
interesting  physical situation which is free of gravity  at  the 
linear  (Newtonian) level. Further it has all but  one  curvatures 
zero, which is also an inveriant for spherical symmetry and it  is 
at  any given radius propotional to curvature of sphere  of  that 
radius [2]. This is why we have termed it as  ``minimally''  curved 
consideration  of  MCS  [3-4]. We have  very  recently  considered 
metric  in the Kerr-Schild (KS) form with a view to find  perfect 
fluid  solutions [5]. It is however well-known that perfect  fluid 
is not compatible with the KS form and hence Senovilla et al [6-
7]  have generalized it by replacing flat metric  by  conformally 
flat. This  would mean that metric can be written as conformal  to 
an  original KS metric, though the base space will not in  general 
be a solution of the Einstein equation. 
\b
\n A metric in KS form is given by 

$$ g_{ij} = \eta_{ij} + 2 H l_i l_j \eqno (1) $$
                     
\n where $H $ is a scalar field and $l_i $ is a null vector relative 
to both $g_{ij} $      
and $\eta_{ij} $. For  spherically  symmetric  vacuum  solution; i.e.  the 
Schwarzschild solution, $H $ satisfies the Laplace equation 
$\bigtriangledown^2 H = 0 $. 
It is rather curious to note that spacetime is not flat unless  $H 
= 0 $; i.e. $ H = const. \not= 0 $, represents a curved spacetime 
 and  in 
particular  MCS. Let us consider the smallest deviation from  flat 
spacetime by taking $H $ constant. Recall that for perfect fluid seed 
metric has to be different from vacuum. So we take $H $ constant  and 
write 

$$ \overline g_{ij} = e^{2U} (\eta_{ij} + 2H l_i l_j), ~ H = const.
\eqno (2) $$
         
\n which  for  spherical symmetry can be brought to  the  orthogonal 
form to read [5]

$$ ds^2 = e^{2U} (-dt^2 + k^2 dr^2 + r^2 d \theta^2 + r^2 sin^2
\theta d \varphi^2). \eqno (3) $$ 

\n Here $U = U(r,t) $ and $k^2 = (1 + 2H)^2/(1 - 2H) $ and the 
base metric is MCS.
\b
\n First,  the  perfect  fluid 
conditions  imply $U = U(r) $, and they yield the general  solution 
for unbounded distribution [5],

$$ e^U = r^{-n} \eqno (4) $$
 
\n with 

$$ 8 \pi \rho = {n(n-2) \over k^2 r^{2(1-n)}},~ \rho {n-2 \over n} p
\eqno (5) $$

\n where $k^2 = 1 + 2n(n-2) $ and removable constants have been  transformed 
away. This  is  the general solution which  represents  isothermal 
fluid  as  it admits a barotropic equation of state  and  density 
falls off as inverse square of the curvature radius, $R = r^{1-n} $.
It is  important  to  note that the metric (3)  admits  this  unique 
isothermal distribution [5]. 
\b
\n Thus  the metric ansatz (3) with the unique solution (4)  is  the 
sufficient condition for isothermal perfect fluid model. We  shall 
now  turn  to the necessary condition.For that we have  to  prove 
that  all  spherically symmetric isothermal fluid  solutions  can 
always be cast in the metric ansatz (3).
Let us consider the general spherically symmetric  metric 
  
$$ ds^2 = -e^{\nu} dt^2 + e^{\lambda} dr^2 + r^2 (d \theta^2 +
sin^2 \theta d \varphi^2) \eqno (6) $$
   
\n where $\lambda $  and $\nu $  are to begin with functions of
 both $r $ and $t $. Now $R_{01} = 0 $ 
 will imply $\lambda = \lambda (r) $. With this we have for perfect fluid 

$$ 8 \pi \rho = {1 \over r^2} [1 + e^{- \lambda} ( r \lambda^{\prime}
- 1)] \eqno (7) $$

$$ 8 \pi p = {1 \over r^2} [- 1 + e^{- \lambda} ( r \nu^{\prime}
+ 1)] \eqno (8) $$

$$ 2 \nu^{\prime \prime} + \nu^{\prime 2} - \lambda^{\prime}
\nu^{\prime} - {2 \over r} (\nu^{\prime} + \lambda^{\prime})
+ {4 \over r^2} (e^{\lambda} - 1) = 0 \eqno (9) $$
 
\n where $\lambda^{\prime} = \partial \lambda/\partial r $.
Now implimenting isothermality; i.e. $\rho = \gamma p \sim r^{-2} $, 
which  will  first  determine $\nu = \nu(r) $, 
and  then  we  obtain  the general solution [8] 

$$ e^{\lambda} = const. = k^2_1,~~ e^{\nu} = r^{-2m}. \eqno (10) $$

\n This belongs to the Tolman class of  solutions [9]. 
Thus  we  have  obtained  the  general  solution  for   unbounded 
spherically  symmetric isothermal fluid and the metric  (6)  will 
read  

$$ ds^2 = r^{-2m} [-dt^2 + k^2_1 r^{2m} dr^2 + r^{2(1+m)}
(d \theta^2 + sin^2 \theta d \varphi^2)]. \eqno (11) $$

\n By  redefining  the  radial coordinate  as $\overline r = 
r^{1+m} $, the  above 
metric on dropping overhead bar  takes the form  

$$ ds^2 = r^{{-2m \over 1 + m}} [-dt^2 + k^2_1 dr^2 + r^2
(d \theta^2 + sin^2 \theta d \varphi^2)] \eqno (12) $$   

\n which is exactly in the required ansatz form (3). 
We have thus proved the theorem: 
   The  necessary  and  sufficient condition  for  a  spherically 
symmetric  spacetime   to represent an isothermal  perfect  fluid 
distribution  without  boundary is that it is  conformal  to  the 
minimally curved spacetime. 
\b
\n It  is   quite  remarkable  that  there  exists  a  one  to   one 
association     between     isothermality     and     conformally  
MCS. Isothermality  of  fluid picks up  uniquely  the  conformally 
MCS. This  is really very interesting because it is a rare
case of a specific physical property singling out a 
geometric property. Like the  isothermal  case, 
stresses  generated by $K $ in MCS also fall off as 
$T^0_0 = T^1_1 \sim r^{-2} $. 
These  stresses have been  identified with the  geometric  string 
dust  [10] as well as with gravitational  monopole  [11]. These 
are   very  exotic  objects  which  are  relevant  in  the   very 
extraordinary setting of the very early Universe. The simplest way 
to  get  to  MCS is that it is a  spacetime  corresponding  to  a 
constant  potential. Recall  that H in (1) satisfies  the  Laplace 
equation.However we confess that its interpretation as  potential 
may  not  be  generally agreed upon. This as  well  as  all  other 
related issues have  been taken up in detail in refs. [3-4]. 
\b
\n Viewing MCS as a small departure (as it differs from flat spacetime 
only  in  tidal  acceleration for transverse  motion)  from  flat 
spacetime, it  is interesting that for its conformal spacetime  it 
does  not permit non-static perfect fluid solution. It admits  the 
general solution having three free parameters, which can represent 
bounded fluid spheres as well as the unbounded cosmological model 
considered above. In contrast conformally flat metric admits  non-
conformally  MCS  is  highly  constrained  and  has  the   unique 
spherically symmetric perfect fluid solution. Even though MCS may
be a small deviation from flat spacetime, their conformal spacetimes
have very different physical properties. 
\b
\n Isothermal fluid structures have been considered in  astrophysics 
for  a  long  time  as  an  equilibrium  approximation  to   more 
complicated systems approaching dynamical relaxed  state. Recently 
the  spacetime (11) has been argued as an ultimate end  state  of 
the  Einstein-deSitter universe [8]. It is envisioned  that  the 
Einstein-deSitter  model asymptotically tends to  an  expansion 
free state, and then it condenses into a stable isothermal
 fluid sphere. 
\b
\n Finally  it  is perhaps for the first time such  a  clear  likage 
between  a  certain physical property of matter  and  a  specific 
geometric property of spacetime has been demonstrated. It is also
for the first time that spacetime conformal to a non-flat spacetime
has been considered.

\vfill\eject
\c{\bf References }
\b

\item{[1]} H W Brinkman  Math. Ann. {\bf 91}, 269 (1924). 
\s
\item{[2]} N Dadhich  Ph.D. thesis (Pune University), (1970).
\s 
\item{[3]}  N  Dadhich  On the  Schwarzschild  field  and a little 
beyond, submitted.
\s 
\item{[4]}  N Dadhich and L K Patel, On the Machian  generalization 
of vacuum spacetimes, to be submitted.
\s 
\item{[5]}  N Dadhich, A conformal mapping and isothermal  perfect 
fluid, submitted.
\s
\item{[6]}  J M M  Senovilla and C F Sopuerta  Class.  Quantum  Grav. 
{\bf 11}, 2073 (1994). 
\s
\item{[7]} J Martin and J M M Senovilla  J. Math. Phys. {\bf 27}, 265 (1986).
\s 
\item{[8]}  W C  Saslaw, S D  Maharaj and N Dadhich, An  isothermal 
universe, submitted. 
\s
\item{[9]} R C Tolman Phys. Rev. {\bf 55}, 365 (1934).
\s 
\item{[10]} P S Letelier  Phys. Rev. {D 20}, 1294 (1979). 
\s
\item{[11]} M Barriola and A Vilenkin Phys. Rev. Lett. {\bf 63}, 341 (1989).

 \bye